\documentclass[sigconf]{acmart}
\usepackage{natbib}
\usepackage{graphicx} % Required for inserting images
\usepackage{soul} % for easy strikeouts/underline etc
\usepackage{eurosym}

\AtBeginDocument{%
  }

\copyrightyear{2025}
\acmYear{2025}
\setcopyright{rightsretained}
\acmConference[FAccT '25]{The 2025 ACM Conference on Fairness, Accountability, and Transparency}{June 23--26, 2025}{Athens, Greece}
\acmBooktitle{The 2025 ACM Conference on Fairness, Accountability, and Transparency (FAccT '25), June 23--26, 2025, Athens, Greece}\acmDOI{10.1145/3715275.3732028}
\acmISBN{979-8-4007-1482-5/2025/06}
\begin{document}

%%
%% The "title" command has an optional parameter,
%% allowing the author to define a "short title" to be used in page headers.
\title{Red Teaming AI Policy: A Taxonomy of Avoision and the EU AI Act}

%%
%% The "author" command and its associated commands are used to define
%% the authors and their affiliations.
%% Of note is the shared affiliation of the first two authors, and the
%% "authornote" and "authornotemark" commands
%% used to denote shared contribution to the research.
\author{Rui-Jie Yew}
\authornote{Both authors contributed equally to this research.}
\affiliation{%
  \institution{Brown University}
  \city{Providence}
  \country{USA}
}
  \email{ryew@cs.brown.edu}
\author{Bill Marino}
\authornotemark[1]

\affiliation{\institution{University of Cambridge}
  \city{Cambridge}
  \country{UK}
}
  \email{wlm27@cam.ac.uk}
\author{Suresh Venkatasubramanian}
\affiliation{%
  \institution{Brown University}
  \city{Providence}
  \country{USA}}
\email{suresh@brown.edu}

%%
%% By default, the full list of authors will be used in the page
%% headers. Often, this list is too long, and will overlap
%% other information printed in the page headers. This command allows
%% the author to define a more concise list
%% of authors' names for this purpose.
\renewcommand{\shortauthors}{Yew*, Marino*, and Venkatasubramanian}

%%
%% The abstract is a short summary of the work to be presented in the
%% article.
\begin{abstract}
 The shape of AI regulation is beginning to emerge, most prominently through the EU AI Act (the ``AIA''). By 2027, the AIA will be in full effect, and firms are starting to adjust their behavior in light of this new law. In this paper, we present a framework and taxonomy for reasoning about ``avoision'' ---conduct that walks the line between legal avoidance and evasion --- that firms might engage in so as to minimize the regulatory burden the AIA poses. We organize these avoision strategies around three ``tiers'' of increasing AIA exposure that regulated entities face depending on: whether their activities are (1) within scope of the AIA, (2) exempted from provisions of the AIA, or are (3) placed in a category with higher regulatory scrutiny. In each of these tiers and for each strategy, we specify the organizational and technological forms through which avoision may manifest. Our goal is to provide an adversarial framework for ``red teaming'' the AIA and AI regulation on the horizon.
\end{abstract}

%%
%% The code below is generated by the tool at http://dl.acm.org/ccs.cfm.
%% Please copy and paste the code instead of the example below.
%%

% %%
% %% Keywords. The author(s) should pick words that accurately describe
% %% the work being presented. Separate the keywords with commas.

% \received{20 February 2007}
% \received[revised]{12 March 2009}
% \received[accepted]{5 June 2009}

%%
%% This command processes the author and affiliation and title
%% information and builds the first part of the formatted document.
\maketitle
\section{Introduction}\label{sec:intro}
As Julie Cohen writes: ``Power interprets regulation as damage, and routes around it''~\cite{schneier2023hacker}. When faced with regulations, firms tend to take the path that minimizes their costs. One way this can manifest is through \textit{avoision}  ~\cite{katz1996ill}: behavior that walks the line between avoidance and evasion by  ``evad[ing] [a] law's intent or purpose'' without it ``actually constitut[ing] unlawful behavior''~\cite{turner2001reactions}. Closely tied to concepts like \textit{regulatory arbitrage}~\cite{pollman2019tech} and \textit{legal hacking}~\cite{alexander2017hacking}, avoision has been examined in the context of copyright~\cite{giblin2014aereo, yim2014normativeavoision}, labor~\cite{turner2001reactions}, and employment law~\cite{alexander2017hacking}. Here, we examine it in the context of artificial intelligence (``AI'') regulation.

In particular, we examine the avoision that firms might engage in as a response to the European Union's Artifical Intelligence Act (the ``AIA''). In doing so, we propose a taxonomy with three tiers of avoisive conduct: (1) the AIA's scope; (2) the AIA's exemptions; or (3) the AIA's consequential categories (of AI, of risk, and of AI operators). 
Within each tier, we hypothesize the technological and organizational embodiments of AIA avoision that may manifest. We justify our hypotheses with either contemporary empirical observations or by pointing to analogous examples of avoisive behavior from other domains. We explain how, although potentially technically permissible under the AIA, these avoisive behaviors can lead to outcomes that run contrary to the AIA's intent --- by, for example, causing harm or undermining fundamental rights in the European Union (EU). To structure our argument, we introduce a taxonomy for identifying and analyzing avoision that may arise in response to the AIA and future AI regulation. We conclude by offering a set of specific policy recommendations for mitigating AIA avoision. Our hope is that the framework we introduce can be used to bring these behaviors to the fore to inform enforcement and reform efforts.

\section{Background}
\subsection{The EU AI Act}

The AIA lays out a uniform legal framework for the development, placing on the market, putting into service, and the use of AI systems in the EU \citep[Rec. 1, Art. 1(2)]{europa}. This piece of legislation went into force on August 1, 2024 and will be in full effect by 2027~\citep{techcrunchEUsGets}. In this section, we lay a foundation for the analysis that follows by examining the history and intent of the AIA, some of its key characteristics, and its cost of compliance (a key motivator for avoision).

\subsubsection{History and Intent}\label{sssec:intent}

Because avoision involves behavior that defies a law's intent, we ground our analysis in the legislative intents that underlie the AIA. They are on display as early as 2018, when the European Commission (EC) proposed that, as part of a broader strategy for AI, the EU develop a regulatory framework for AI that would simultaneously ``promote innovation while ensuring high levels of protection and safety'' \citep{ec2018coordinated}. Despite the ``lengthy legislative process'' \citep{lomas2024eu} and ``heated debates''\citep{kelder2024relative} that followed, this ``dual purpose'' \citep{laux-wachter} remained largely intact through the AIA's passage. As stated in its Recitals, the AIA aims to simultaneously foster innovation and promote the uptake of ``human centric and trustworthy artificial intelligence'' while also protecting health, safety, and the fundamental rights encoded in the Charter of Fundamental Rights of the European Union~\citep{eu_charter_2000} and guarding against any ``harmful effects'' AI could have in the EU~\citep[Rec. 1]{europa}. 

In pursing the twin objectives of innovation and protecting fundamental rights, the AIA targets some sub-goals that are themselves worth highlighting for our analysis. First, the AIA routinely aims its rules at ``ensur[ing] a level playing field'' for AI providers~\citep[Rec. 8]{europa}, with some rules customizable to ``the size of the provider'' to avoid putting undue burdens on small and medium enterprises (SMEs) and startups~\citep[Rec. 109]{europa}. This has led some commentators to view the AIA as ``primarily target[ing] the largest tech platforms capable of the greatest harms while also seeking to level the playing field so new companies can enter the market without being swallowed up''~\citep{hodge2022experts}. Secondly, because the AIA's aim is to ``ensure an effective protection of natural persons located in the Union'' the Recitals also explain that the AIA sometimes applies to entities that are established outside of the EU~\citep[Rec. 22]{europa}. When it comes to the ``effective protection of rights and freedoms of individuals across the Union,'' the AIA wants AI providers based in or outside the EU to be on the same footing~\citep[Rec. 21, 22]{europa}. This intended ``extraterritorial effect'' \citep{Bellogin2024EUAIACT} explains, among other things, the need for providers established in third countries to appoint an authorized representative established in the EU \citep[Rec. 82]{europa}.\footnote{Some see these provisions as an attempt ``to eliminate
opportunities for regulatory arbitrage'' by preventing AI developers from changing the locations of themselves or the AI systems and models to avoid the AIA \citep{lancieri_ai_regulation_2024}. Later in this paper, we argue that similar opportunities can still be had despite these provisions.}

\subsubsection{Key Features}\label{ssec:definingfeatures}

The text of the AIA has some defining elements, many of which are spawned by the intents described in Section~\ref{sssec:intent}. These features, in turn, play an important role in shaping our analysis--in particular, the tiers of avoision in our taxonomy. One important product of the AIA's twin objectives of ``minimiz[ing] threats while promoting innovation and market diffusion''\citep{wahlster2022germanroadmap} --- and its ``defining feature'' \citep{ustaran2024top} --- is its ``risk-based'' approach to regulation \citep{Mahler2022-gc}. Under this approach, the AIA ``tailor[s] the type and content of such rules to the intensity and scope of the risks that AI systems can generate''  \citep[Rec. 26]{europa}. In practical terms, what this means is that the AIA's ``strictest''\citep{weatherbed2023euai} requirements are reserved for AI systems deemed to be ``high-risk''~\citep[Art.6; Sec. 2]{europa} and those GPAI models deemed to carry ``systematic risk''~\citep[Art. 55]{europa}.

Cognizant that parties further upstream in the AI supply chain may be in a better position to mitigate harms~\citep[Rec. 101]{europa}, the AIA defines several AI ``operator'' roles spanning the AI value chain ---including providers, deployers, importers, and distributors--- and assigns varying obligations for each \citep[Art. 3(8), 25]{europa}. Among these, the ``provider'', defined as anyone that develops an AI project (or that has an AI project developed) and places it on the market or puts it into service under their own name or trademark \citep[Art. 3(3)]{europa}, tends to ``bear the largest share of obligations'' \citep{cdtPublicAuthorities}. However, in certain scenarios that we describe later in our paper, other operator roles can assume the responsibilities normally placed on the provider \citep[Art. 25]{europa}.

Finally, the AIA wholly or partially exempts multiple types of AI systems and GPAI models, usually to advance the goals of fostering innovation discussed in \ref{sssec:intent}. For example, the AIA contains exceptions for open source efforts \citep[Arts. 2(12), 53(2)]{europa} that, it says, tap into open source's ability to ``contribute to research and innovation in the market'' --- thereby aligning with the AIA's stated goal of fostering innovation \citep[Rec. 102]{europa}. Likewise, the AIA contains exemptions for AI systems and GPAI models that are ``developed and put into service for the sole purpose of scientific research and development''~\citep[Art. 2.6]{europa} --- another measure intended to ``support innovation''~\citep[Rec. 25]{europa}.

\subsubsection{Cost of AIA Compliance}
\label{sec:cost}

It has been said that the ``costs associated with [AIA] compliance will be substantial'' \citep{hacker2024comments}. This is because compliance entails not only the work of understanding the AIA, which may require legal or compliance expertise, but also the subsequent development work of integrating compliance measures into the AI system or model \citep{mitOrganizationsFace}. As a result, bringing some AI systems and models into compliance with the AIA has been estimated to cost as much as \euro{400,000} \citep{europa2, koh2024voices}. Put differently, it is expected that the AIA will add an overhead of as much as $17\%$ to the development costs of affected AI efforts \citep{laurer2021clarifying}. These high costs create economic incentives for AI firms to circumvent as much of the AIA's requirements as possible. While doing so, however, they are also incentivized, by the AIA's heavy penalties (of up to \euro{35,000,000} or 7\% of global revenue \citep[Art. 99(3)]{europa}), to comply with the letter of the law. This can serve to explain the potential attraction of \textit{avoision} as regards the AIA.

\subsection{Related Work}
Our work, while focusing specifically on the AIA, embodies the spirit of literature that applies the adversarial mindset of computer security and systems to policymaking and compliance. This literature includes \citep{chilson2024red}, which considers ``red-teaming'' California's SB 1047. Red teaming is a concept originating in computer security and systems, but has gained increasing traction in the context of AI.\footnote{It is defined within the US Executive Order on the Safe, Secure, and Trustworthy Development and Use of Artificial Intelligence as  ``a structured testing
effort to find flaws and vulnerabilities in an AI system,
often in a controlled environment and in collaboration
with developers of AI''~\citep{eo2023safe, feffer2024red}.}
\citet{schneier2023hacker} similarly draws an analogy between ``vulnerabilities'' in computer code and ``loopholes'' in legal codes. \citet{schneier2023hacker} additionally discusses the industry of tax attorneys and accountants as an industry of ``black-hat'' hackers, who intend on finding loopholes in the tax code. In this paper, we adopt this adversarial mindset in our structured framework for reasoning through the organizational and technological maneuvers that may be leveraged to subvert the AIA. 

Also relevant to the topic at hand, other research efforts have highlighted the AIA's loopholes and the opportunities for regulatory arbitrage that it presents. For example, \citet{wachter26limitations} analyzes some of its ``many problematic ambiguities, enforcement gaps, and loopholes'', critiquing, among other things, its self-assessed filter provision ~\citep[pg. 693]{wachter26limitations} (a point that is echoed in other scholarly works \citep{hacker2024comments, smuha2021eu}). Outside of work by \citet{wachter26limitations}, \citet{colonna2023ai} investigates opportunities for regulatory arbitrage via the AIA's research exemption. Meanwhile, other work points to the compute thresholds referenced in the definition of GPAI models as a gameable proxy measure insufficient for robust AI governance~\citep{hooker2024limitations}. While our work implicates some of these loopholes and enforcement gaps in the AIA and other AI regulations, our main contributions lie in the framework we introduce and on the specific strategies that firms might undertake in circumventing the AIA. 
 In work aimed not at the AIA but at the EU's Artificial Intelligence Liability Directive \citep{EU2022AILD},  \citet{wachter26limitations} puts forth the idea of firms adding a ``token `human in the loop''' to AI systems to avoid liability ~\citep[pg. 693]{wachter26limitations}. \citet{crootof2023humans} similarly discusses how a human-in-the-loop can arbitrage software definitions as part of the U.S. ``FDA Cures Act''. In this piece, we extend this concept of a human-in-the-loop to avoision under the AIA.

\section{Methodology}
As described in Section~\ref{sec:intro}, this paper presents a framework and taxonomy for reasoning about avoision in response to the AIA. In doing so, we hypothesize a number of avoisive behaviors that we group into three distinct tiers. In order to ensure that these various behaviors are not only plausible but adhere to definitions of avoision articulated by other scholars \citep{katz1996ill, turner2001reactions, giblin2014aereo}, we include only behaviors that:
\begin{enumerate}
    \item Are not unlawful evasion of the AIA, but, rather, comply with the letter of the law; 
    \item Despite complying with the letter of the law, run contrary to the intent of the AIA in one way or another;
    \item Are economically rational to pursue (e.g., offer some tangible benefit compared to compliance, such as reduced compliance costs)
\end{enumerate}

We point to contemporary, empirical observations to support these hypothesized behaviors. However, because the AIA is still relatively new and firms are still figuring out their response to it, there is (so far) little documentation of avoision in the wild as regards the AIA. As such, we often point to historical precedent to bolster our hypotheses: that is to say, past avoision from analogous areas of law and technology. While these earlier episodes may be drawn from non-EU jurisdictions, they are still relevant precedent for this analysis because they demonstrate how economically rational firms could theoretically respond to similar scenarios brought about by the AIA.

\section{A Taxonomy of Avoision}
In this section, we articulate our three-tiered taxonomy of AIA avoision. The first captures avoisive behavior targeting the AIA's \textit{scope}. The second captures avoisive behavior targeting the AIA's \textit{exemptions}. Finally, the third captures avoisive behavior targeting the AIA's \textit{consequential categories}, which affect the regulatory burden an AI system must bear: that is, its categories of AI, categories of risk, or categories of AI operators. With each subsequent tier in our taxonomy comes a higher level of regulatory burden on the avoisive firm in terms of the severity of  the rules that need to be considered. Behavior in the first tier aims to avoid the regulatory burden of the AIA altogether by circumventing its scope of application. Behavior in the second tier aims to avoid all or some of the AIA's requirements by fitting itself into one of the AIA's whole or partial exemptions. Behavior in the last tier, meanwhile, only aims to reduce (not eliminate) the AIA's regulatory burden by fitting itself into a less-regulated category. A more detailed description of each of these three tiers and their associated strategies can be found in Table~\ref{table:taxonomy}.

\begin{table*}
\caption{\label{table:taxonomy} We define components of our taxonomy and consider how responses to a variety of regulations in the past fit within the taxonomy we present.}
\begin{tabular}{p{6cm} p{9.5cm}}
\toprule
\textbf{Tiers}&\textbf{Historical Examples}\\\hline\\
\textbf{Scope}: This first tier captures maneuvers to avoid the AIA completely by rendering AI systems or models out of \textbf{\emph{scope}}. Within this tier, we consider avoision strategies centered on how AI companies might structure their AI systems, models, or outputs in relation to the EU market.  &  When the EU began regulating genetically modified organisms (GMO), seed producers used nuclear radiation and mutagenic chemicals to produce unregulated crops --- crops that were technically not ``genetically modified'' --- which nonetheless had similar qualities to GMO seeds~\citep{burk2016perverse}. By using these other techniques, developers moved their activities out of scope of the regulation--perhaps, in opposition to the regulation's intent to manage harms arising from the use of GMOs.
\\\\
\textbf{Exemptions}: This second tier encompasses strategies that can be leveraged to take advantage of total \textbf{\emph{exemptions}} in the AIA. Within this tier, we consider how actors might take advantage of exemptions for open-source models and models that are released for the sole purpose of scientific research in ways that are anti-competitive and keep key details behind closed-doors. &  \citet{schneier2012liars} highlights an instance in which Japan routinely exploited a 'scientific research' exemption in the International Whaling Commission’s whale hunting regulations to ``kill[] about 330 whales each year for `research' ''. A more banal example involves US patent law's ``experimental use exemption'', a court-created defense to patent infringement through which multiple defendants have (unsuccessfully) pursued revenue-enhancing activities ~\citep{Mueller2005TheEE, 890b3fb3-6483-3ede-89ea-004f2e17c043}. 
\\\\
\textbf{Categories}: This final tier considers the avoidance of consequential \textbf{\emph{categories}} under the AIA. The AIA places higher compliance burdens on some types of AI systems and models (e.g., high-risk AI systems) and some types of AI actors (e.g., providers or developers). Within this tier, we consider how AI actors might toe the line to avoid burdensome role requirements and between different types of AI in relation to the AIA. & Against the backdrop of burdensome requirements set forth by the United States Environmental Protection Agency (EPA), Chrysler designed its 2001 mini-cruiser vehicle such that the the distance between the wheels of the vehicle allowed the company to avoid the categorization of ``passenger cars'' and fall into the categorization of ``light trucks''~\cite{burk2016perverse}. However, light trucks were also subject to a tax for importation. To deal with international tax laws, the company imported its vehicles first as ``passenger cars'' to enjoy a tax break and then tore out its seats so that its vehicles could again be categorized as ``light trucks''.  The company was then able to avoid the tax and benefit from the EPA's less burdensome requirements by first importing a given vehicle as a ``passenger vehicle'' and then removing the rear seats and seatbelts of the vehicle so that it could be classified once again as a ``light truck''.\\
\end{tabular}
\end{table*}

\subsection{Avoision Targeting the AIA's Scope}\label{sec:scope}
Whether due to its discrete objectives or the particular debates surrounding its passage, the AIA's scope (laid out in ~\citep[Art. 2]{europa}) is decidedly ``limited''~\citep{villegas2024us}. For example, this scope is restricted to ``artificial intelligence''~\citep[Art. 1]{europa}, with the majority of the AIA's rules applying only to two neatly-defined categories of AI: AI systems~\citep[Art. 3(1)]{europa} and general-purpose AI (GPAI) models~\citep[Art. 3(63)]{europa}.  Within these two categories, the AIA's scope is further limited to only those systems and models connected to the EU market in one of the ways laid out in ~\cite[Art. 2]{europa}: i.e., those that are ``placed on the market'' or ``put into service'' in the EU, have their ``outputs'' used in the EU, or are deployed by an EU-based deployer ~\citep[Art. 2.1(a-c)]{europa}. Presumably, firms who can position their technologies outside of these discrete settings will entirely avoid having to comply with the AIA's requirements --- no small matter given the high cost of AIA compliance described in Section~\ref{sec:cost}. Here, we map out how firms might try to accomplish this maneuver through two exemplary strategies: (1) circumventing the AIA's definitions of AI systems; and (2) distancing their AI systems and models' from the EU market.

\subsubsection{Circumventing the AIA's Definition of AI Systems}\label{sssec:aidefs}
Although the AIA's scope is ostensibly restricted to ``artificial intelligence,'' it does not define the term  ~\citep[Art. 1]{europa}. That said, it does define the discrete categories of AI that the majority of its provisions regulate: AI systems and GPAI models. When it comes to AI systems~\citep[Art. 3(1)]{europa}, the AIA defines these to\ include any ``machine-based system that is designed to operate with varying levels of autonomy and that may exhibit adaptiveness after deployment, and that, for explicit or implicit objectives, infers, from the input it receives, how to generate outputs such as predictions, content, recommendations, or decisions that can influence physical or virtual environments''~\citep[Art. 3.1]{europa}. This definition has been called ``[o]ne of the most consequential concepts'' in the AIA because it ``determines which systems fall within the scope of the regulation and may therefore be subject to [its] requirements''~\citep{Fernández-Llorca2024}. While some scholars have called this definition ``broad'' \citep{grady2023ai, ruschemeier2023ai, Quattrocolo_Sacchetto_2024, madiega2024briefing}, we argue that the definition may be discrete enough to create openings for at least two types of avoisive behaviors. The first of these targets the fact that the definition of AI systems only includes those that are ``machine-based'' and ``designed to operate with varying levels of autonomy'' \citep[Art. 3.1]{europa}. The second, meanwhile, targets the exceptions to the definitions that are explicitly stated in the AIA's Recitals \citep[Art. 12]{europa}. 

\paragraph{Adding Human Veneers to Make AI systems Less ``Machine-based'' or ``Autonom[ous]''} Under the AIA's definition, AI systems must be ``machine-based'' and be ``designed to operate with varying levels of autonomy'' \citep[Art. 3.1]{europa}. The AIA clarifies that ``machine-based'' means ``that AI systems run on machines'' \citep[Rec. 12]{europa}. ``Autonomy,'' meanwhile, means the AI system has ``some degree of independence of actions from human involvement and of capabilities to operate without human intervention'' \citep[Rec. 12]{europa}. Put together, these two requirements appear to create a loophole for systems that are not run entirely on machines (i.e., run partially ``on humans''), that display no real autonomy, or both. For example, a growing trend involves AI models providing ignorable ``advice'' to humans as they make decisions or perform tasks \citep{jourdan2024dissertation, chong2022human}. These systems only run partially on machines and, given that they cannot take any action without human input \citep{hauptman2024understanding}, display no real autonomy. Accordingly, there is a valid question about whether they are AI systems under the AIA's definition.  Firms could foreseeably exploit this loophole by adding one or more human layers --- or \textit{veneers}, if you will --- atop their AI systems, arguably putting them outside the scope of the AIA.\footnote{Note that this idea of adding a human ``veneer'' is spiritually similar to the idea, put forth in \citep{wachter26limitations}, that firm might add a ``token `human in the loop''' to AI systems to try to avoid liability under the EU's Artificial Intelligence Liability Directive.} To draw an analogy from the U.S. regulatory landscape, it has been argued that regulation on medical devices can sometimes be circumvented by positioning a human medical professional reviewer at the output end of these technologies~\citep{crootof2023humans}. 
    % As an example, firms could shift from fully AI-based loan approval systems to systems that merely use AI to generate reports that a human loan officer optionally consults before deciding whether or not to make a loan. 
    The danger here --- and the reason this practice represents intent-undermining avoision --- is that adding these veneers may not reduce the risk of harm posed by the underlying AI. Humans tend to over-rely on AI advice, often perpetuating its inaccuracies and biases \citep{passi2022overreliance}.\footnote{Note that, while precautions --- including some of the human oversight tools prescribed in Article 14 of the AIA \citep[Art. 14]{europa} --- can theoretically reduce this risk, they would not be obligatory if the AI system is outside the AIA's scope.} Therefore, these human-veneer architectures, despite potentially complying with the letter of the law, would likely undermine the AIA's intent of preventing harm.
    
    \paragraph{Adding Rule-based or Traditional Software Veneers to AI Systems (i.e., ``Reverse AI-washing'')} The AIA also explicitly states that its definition of AI systems does not include ``simpler traditional software systems or programming approaches'' or ``systems that are based on the rules defined solely by natural persons to automatically execute operations''~\citep[Rec. 12]{europa}. With regard to the latter, it has been said there is no ``clear red line'' about what is or isn't excepted~\citep{hacker2024comments}. Firms could foreseeably exploit these exceptions --- or the ambiguity enshrouding them --- through what we playfully deem ``reverse AI washing''\footnote{``AI-washing'' captures the phenomenon where business incentives encourage technologies underpinned by human labor or logic-based algorithms to be marketed as AI systems~\citep{woollacott2024ai}. It therefore seems appropriate to call the near-inverse of this --- positioning AI systems as non-AI in order to achieve regulatory objectives --- ``reverse AI washing''.}: adding one or more veneers of rule-based or traditional software atop an AI system in order to wedge it into one of these stated exceptions to the defition of AI systems. For example, AI can be used to pre-process input data for downstream rule-based inference \citep{KIERNER2023104428}. Differently, AI can be used to generate static data that is indexed before being accessed downstream by traditional software algorithms like keyword-based search \citep{zhang2024valueaigeneratedmetadataugc}. Importantly, such fusion architectures may not ultimately reduce the risks of harm when compared to AI systems; AI risks would instead be passed downstream to the wrapping rules-based or traditional software algorithm, thereby potentially sabotaging the AIA's intent of reducing harm.

\subsubsection{Distancing AI Systems --- But Not Other Elements of the Business --- from the EU Market}

Non-EU-based providers of AI systems that are not \textit{placed on the market} in the EU, are not \textit{put it into service} in the EU, and whose outputs are not \textit{used} in the EU are outside the scope of the AIA ~\citep[Art. 2.1(a-c)]{europa}. AI providers are incentivized to fit themselves into these criteria, since doing so will mean shrugging off the AIA's regulatory burden entirely. On the other hand, AI providers may be loathe to give up the revenue generated in the EU, ``one of the world’s biggest markets'' ~\citep{wachter26limitations}. Accordingly, what we foresee are strains of avoision whereby AI providers seek to carefully organize their AI systems and deployments so as to fit themselves inside the criteria above, putting themselves out of the scope of the AIA, while also benefiting, in one way or another, from the EU market or EU users. To provide examples of this strain of avoision, in this section we consider how the use of non-AI wrappers (both human and technological) and the strategic geographical placement of components of the AI systems may be leveraged to create a plausible argument that AI system is not ``plac[ed] on the market" in the EU'', ``ma[de] available...on the Union market''~\citep[Art. 3.9]{europa}, or ``used in the Union''~\citep[Art. 2.1(c)]{europa}, and is therefore outside the AIA's scope.

\paragraph{Human AI Wrappers}

AI systems could be configured such that all AI models run outside of the EU and, furthermore, their outputs are delivered initially to non-EU-based humans that act as a ``wrapper'' on those AI outputs, using them to inform a human decision-making process which, in turn,  impacts EU users. Thus, AI ``outputs'' may not considered to be used in the EU market ~\citep[Art. 2.1(c)]{europa} and, arguably, are therefore outside the scope of the AIA. By way of background, when the provider or deployer of the AI system is in a third country, the AIA kicks in ``where the output produced by the AI system is used in the Union'' ~\citep[Art. 2.1(c)]{europa}. 
 The insertion of a human decision-maker in another country potentially creates opportunity for an argument that it is the human's decision that ultimately touches the EU market, not the AI output.\footnote{We do note, however, that the AIA's drafters had some degree of foresight regarding this case. Namely, \cite[Recital 22]{europa} states: ``To prevent the circumvention of this Regulation and to ensure an effective protection of natural persons located in the Union, this Regulation should also apply to providers and deployers of AI systems that are established in a third country, to the extent the output produced by those systems is intended to be used in the Union''. The recital further states that its motivation for doing so is to ``prevent the circumvention of this Regulation''. However, in this example, the insertion of a human decision-maker in between an AI decision could also challenge the extent to which the output used downstream in the EU is indeedproduced by an AI system (rather than a human one).}

\paragraph{Technological Non-AI Wrappers}\label{par:techwrap}
Non-AI technologies  deployed outside of the EU could also be used to wrap the outputs of AI systems in ways that make it less clear whether the AI is being used in the EU~\citep[Art. 2.1(c)]{europa} and is therefore within the scope of the AIA. As an example, consider an AI system that inputs a video into an AI system that leverage different AI models (e.g., speech, visual, and audio classifiers) to produce labels or tags describing the content in each of these different modalities. This mixture of tags is then transmitted to a separate,  
 non-AI (e.g., rule-based) system that harmonizes those various tags into a single label or tag for the video that can be used to enhance video search experiences. The AI system and its component models as well as the non-AI system can be placed outside of the EU, with only the final tag outputted by the non-AI system being transmitted into the EU for use during search by EU users. In this case, it arguably becomes less clear whether ``the outputs of the AI system are ... used in Union''~\cite[Art. 2.1(c)] {europa}. The point at which a system or algorithm should be considered AI is a point of current contention.\footnote{Workday, a job recommendation engine and recruiting platform, is proposing AI policies that state that AI that falls under the scope of the policy must be a ``controlling'' factor in a ``consequential decision''~\citep{workday2024model}. This potentially gives Workday room to argue that the process of collating AI signals no longer subjects its systems to the regulations it is proposing.}

\paragraph{Server Placement and Data Storage} 
Given that the AIA applies only to deployers that have their place of establishment in the Union~\citep[Art. 2.1(b)]{europa}, deployers may choose to place servers or run computations out of the EU. Using server location to manage liability under Internet laws has a long history.\footnote{When the EU Privacy Directive was put forth in 1998, Robert E. Litan provided testimony that: ``...it is hard to believe that any government agency charged with enforcing a broad privacy law could ever stay ahead of the game, let alone discipline operators of sites who move their servers off-shore''~\cite{litan1998european}. Similarly, as part of a discussion on how to regulate the Internet in the United Kingdom,  Dr. Konstantinos Komaitis noted that burdensome UK regulations  ``will serve to trigger a shift of online platform hosting providers from having their \textbf{servers based in the UK to migrate them abroad} to more lenient regulation regimes''~\citep{isuk2018internet}. As another example, in light of the U.S. Section 230, lawyers in Italy and South Korea have expressed their lack of leverage to pressure U.S.-hosted sites to take down their citizens' content because they do not have jurisdiction over them~\cite{citron2023fix}.} Under data protection and privacy laws, AI companies have already invoked server placement to limit liability. 
In court for alleged violations of the Illinois Biometric Information Privacy Act (BIPA), Google leveraged the extraterritoriality doctrine to argue that, because its conduct occurred in the cloud and on remote servers, the company should not be subject to BIPA.\footnote{However, the District Court denied Google's motion to dismiss, focusing instead on the fact that the photographs from which biometric identifiers were extracted were taken by Illinois residents and uploaded through an Illinois IP address.} The AIA is a cross-jurisidictional regime\footnote{``[T]he EU AI Act is a cross-jurisdictional regulatory regime where providers of AI systems may be subject to its rules as long as there is an impact on the EU market, independent of where providers' are physically located'', something which may represent "a push to eliminate opportunities for regulatory arbitrage''~\citep{lancieri_ai_regulation_2024}}, but there may still be incentives for firms to move training activities outside of the EU under the AIA~\citep{lancieri_ai_regulation_2024} . \citep[Art. 53(1)(c)]{europa} requires that providers ``put in place a policy to comply with Union law on copyright and related rights''. However, whether EU copyright law ``actually covers machine learning activities that take place outside the EU''~\citep{lancieri_ai_regulation_2024} remains an open question. By moving AI training (the processing of copyrighted works) outside the EU, firms could take advantage of ambiguity in EU copyright law's extraterritoriality principle to limit the extent to which the AIA applies. The resulting AI models and systems may nonetheless impact the EU market.

\subsection{Avoision Targeting the AIA's Exemptions}\label{sec:exemptions}
This next tier of avoision strategies captures how firms might try to avoid the compliance burden imposed by the AIA --- not by positioning themselves \emph{outside} its scope, but by positioning themselves \emph{inside} its exemptions. In particular, firms may seek to exploit carve-outs the AIA offers for scientific research~\citep[Art. 2.6]{europa} and open-source AI systems and models ~\citep[Art. 2.12, 53.2]{europa}.  A firm that seeks to take advantage of AIA exemption articles through open-sourcing or the promotion of scientific research could be doing this exactly in alignment with the AIA. However, by doing so without promoting competition or openness of findings as the AIA intends, firms may confer benefits to their organization such as talent and intellectual property as we detail below --- while disencumbering themselves of AIA requirements and in misalignment with the AIA's intent.

\subsubsection{Exploiting the Research Exemption}
 The AIA contains a total exemption for AI systems or models ``specifically developed and put into service for the sole purpose of scientific research and development''~\cite[Art. 2.6]{europa}. This exemption does not clearly exclude commercial research and development (``R\&D'') activities \citep{colonna2023ai}. Accordingly, firms could seek to ``circumvent AI Act requirements by structuring work as R\&D'' \citep{bogucki}, violating the intent of the exemption (or the AIA overall) as they do. As also detailed in Section~\ref{sssec:intent}, AIA Recitals state that the Act should ``support innovation, should respect freedom of science, and should not undermine research and development activity''~\cite[Rec. 25]{europa}. As a whole, the intent of the AIA includes supporting SMEs, like start-ups~\citep[Rec. 8]{europa}. In this section, we consider how avoision to the research exemption might instead close off knowledge, undermine competition, and sidestep regulation of commercial AI investments in contravention of the AIA's goals.\footnote{A similar point has been made before. In particular, the research exemption creates a ``potential loophole which makes it possible for providers and users of systems which would otherwise be subject to the requirements for high-risk or prohibited systems to claim that they are merely doing ‘in-the-wild' experiments, rather than actually developing or deploying a high-risk or prohibited AI system.'' \citep{smuha2021eu}}

 \paragraph{Research without Openness}
 Although the AIA itself does not explicitly tie this exemption to openness of scientific research, some scholars have suggested that, given the ``EU’s push towards an open science model'', it would be ``inequitable for ... entities to invoke the research exemption in situations where their research results will not be made publicly available''
 \citep{colonna2023ai}.\footnote{It would also clash with the principle of ``strengthening [the EU's] scientific and technological bases by achieving a European research area in which ... scientific knowledge and technology circulate freely'' \citep[Article 179(1)]{europatfeu}, which is not only encoded in the EU's founding charter but was a stated goal of the GDPR's similar research exemption \citep[Art. 9(2)(j), 89]{europagdpr}.} Therefore, some believe that the exemption carries an ``ethical obligation to make [research] results open and publicly accessible''~\citep{quezada}. AI firms could therefore be undermining the goals of the Act by using the research exemption to sidestep the AIA and incurring the various internal benefits of investing in R\&D \citep{tsai, abed170b-70e1-3d1a-8ab9-48fe7e80ae11, ANZOLAROMAN2018233, worksinprogressRiseFall} without sharing any valuable scientific knowledge back to the larger EU AI community. One way this could happen is if the firms that leverage the exemption do not publish their research findings or publish them in a limited or hard-to-reproduce manner (e.g., without code or documentation) \citep{zdnetState2020, Gundersen_2020, Haibe-Kains2020}. Differently, those firms could publish scientific research which is valueless to the larger EU market because they alone are in a position to internalize its benefits --- e.g., because they have siloed the hardware or engineering talent needed to productionize the research~\citep{jurowetzki2021privatizationairesearcherscauses}).
 
\paragraph{Research to Control and Restrict} Firms may restructure market-targeting work as R\&D to sequester valuable resources such as talent and intellectual property. Because ``[the] ability to do creative, `blue-skies' research and gain academic esteem'' is a known draw for AI talent~\citep{jurowetzki2021privatizationairesearcherscauses}, firms could leverage the exception to keep top researchers siloed and away from startups, competitors, and academia. By patenting or otherwise protecting the outputs of R\&D, firms can create so-called ``patent thickets'' that deter the entry of competitors into a space \citep{a8408bc4-9f57-321e-8b14-2905464a27af}, even if they themselves are not yet mining it commercially. Once talent has been cornered and the market has been flushed of competitors, the firm, in a stronger position to absorb the cost of regulation, could commercialize the research and waive the exemption. While these practices may not violate the letter of the law, they would undermine the spirit of the AIA, which, in multiple places, recites its goals of ``ensur[ing] a level playing field'' for AI operators  \citep[Rec. 21, 82, 106]{europa} and fostering ``competitiveness'' in the EU market \citep[Rec. 121, Art. 40(3), Art. 57(9)]{europa}, instead exacerbating the concentration of resources already on display in the tech industry \citep{Verdegem2024}.

\paragraph{Willful Blindness as R\&D ``Leaks'' into Commercialization}
It has been said that, in the AI industry, there is a  ``blurry line between research and application'' \citep{scaleEnterpriseStrategy}, with the same team inside a firm sometimes playing both roles \citep{10.1145/2209249.2209262}. When it comes to datasets, it is known that research-only datasets sometimes find their way to commercial use \citep{longpre2023data}, including intentionally through a practice known as ``data laundering'' \cite{waxyDataLaundering}. This may also happen unintentionally \citep{10.1145/3458723}, perhaps in part because data provenance inside data-rich organizations is challenging \citep{dataprovenance, ibmWhatData}. Firms could take advantage of this context to move work into  R\&D, liberating themselves from the requirements of the AIA, while knowing there is some probability the output of that R\&D, including but not limited to datasets, could ``leak'' into commercial projects. This willful blindness could violate the spirit of the AIA as unregulated data or models crossover into initiatives that do reach EU consumers.

\subsubsection{Exploiting the Open-source Exemption}\label{sec:opensourceexploit}

The AIA contains generous exemptions for open-source AI systems and models. Specifically, it offers a total exemption for AI systems ``released under free and open-source licenses''~\cite[Article 2(12)]{europa}.\footnote{This exemption does not apply to AI systems that are high-risk, involve prohibited practices, or are subject to additional disclosure requirements under Article 50 \cite[Article 2(12)]{europa}.} It also offers a partial exemption, waiving some of the AIA's requirements,\footnote{Note that the requirements in \citep[Article 53(1)(c), 53(1)(d)]{europa} are not waived, which is what makes this a partial exemption.} for GPAI models which are ``released under a free and open-source licence that allows for the access, usage, modification, and distribution'' and whose ``parameters, including the weights, the information on the model architecture, and the information on model usage, are made publicly available'' \citep[Art. 53(2)]{europa}.\footnote{Notably, this partial exemption does not apply to GPAI models with systemic risk \citep[Art. 53(2)]{europa}.} As with the scientific research exemption, firms could seek to circumvent AI Act requirements by restructuring their work as open-source projects, reaping various benefits while violating the intent of the exemption (or the AIA overall).\footnote{Note that, on top of all this, these unregulated open-sourced models could carry negative externalities for the EU (e.g., helping propagate models that enable the generation of hate speech or misinformation \citep{theparliamentmagazineOpedEUs}.} 

\paragraph{Open-washing}

The AIA is clear about the intent behind its open source exemption. Open source AI, declares its Recitals, ``can contribute to research and innovation in the market and...provide significant growth opportunities for the [EU] economy'' \citep[Rec. 102]{europa}. In other words, these exemptions are primarily driven by their ``economic benefits'' to the EU \citep{DBLP:journals/corr/abs-2405-08597}. For these benefits to accrue, however, it has been argued that firms who take advantage of the exemptions must ``actually contribut[e] to the commons'' \citep{liesenfeld2024rethinking} --- i.e., they must open-source something that is valuable to the EU's AI ecosystem and economy at large. But the fact that ``the Act doesn’t actually bother to define exactly what it means for models to be under `free and open source licenses' '' \citep{downing} creates an opportunity for firms to exploit this exemption, avoiding regulation while reaping the strategic benefits of open-source \citep{wiredWIREDGuide, DBLP:journals/computer/Riehle07, jdsupraActOpenSource, https://doi.org/10.3401/poms.1070.0004, doi:10.1080/07421222.2014.995564} and giving little of value to the ``commons.'' This practice has been called
``open-washing''~\citep{widder2023open} and may include withholding elements of the technology itself, withholding documentation, or limiting access through sign-up forms or restrictive licenses \citep{liesenfeld2024rethinking, marbleLicensesMasquerading}. Other strategies yet to be studied might involve open-sourcing models whose inference costs \citep{semianalysisInferenceCost} or specialized hardware requirements \citep{semianalysisInferenceCost, widder2023open} are prohibitively expensive for most market participants (including competitors) or open-sourcing models without critical accompaniments such as safety classifiers, watermarking systems, or explainability tools, rendering the models virtually unusable on the EU market (especially in light of the AIA). All of these techniques undermine the \emph{quid pro quo} implicit in the exemption.

\paragraph{Open-source to Control and Restrict}

Like scientific research, open-sourcing could be misappropriated to make a ``level playing field''~\citep[Rec. 21, 82, 106]{europa} uneven and to reduce ``competitiveness'' \citep[Rec. 121, Art. 40(3), Art. 57(9)]{europa}, undermining stated goals of the AIA. Like open-sky scientific research, the prospect of open sourcing one's research may be a lever that firms can pull to silo top AI talent \citep{jdsupraActOpenSource}. Commentators have also argued that firms could also see open source as an opportunity to ``[o]wn[] the ecosystem'' and ``garner[] an entire planet's worth of free labor''~\citep{semianalysisGoogleHave} as that ecosystem uses an open-source architecture the firm has pioneered to train new models that are easily incorporated into its products~\citep{semianalysisGoogleHave, wsj}. Lastly, large firms may also see open sourcing (and thus commoditizing) AI as a way to neutralize any threat it poses to their market dominance --- a practice sometimes called ``AI dumping'' \citep{andrewzuoDumping}. What all of this means is that that open-source exception may be a sort of subsidy for anti-competitive behavior, liberating it from regulation even as it undermines the AIA's own goal of ``enabl[ing] competition and innovation by new entrants and smaller players, including in the EU'' \citep{surman2023openness}.

\subsection{Avoision Targeting the AIA's Consequential Categories}\label{sec:categorizations}

Under the AIA, the exact requirements that AI must satisfy are a function of the type of AI at issue (whether AI system or GPAI model), its category of risk, and the role of its operator. We call these categories the AIA's consequential categories and posit that avoison will often strategically target classification of an AI system or model (or its operator) in particular categories in order to reduce the regulatory burden posed by the AIA. Avoision in this tier often undermines AIA intent because these strategies can result in categories of AI or AI operators which carry more substantial risk of harm, while avoiding the higher levels of regulation appropriate to their actual risk. In this section, we explore example avoision strategies that target each of the three consequential categories: type of AI, category of risk, and category of operator.

\subsubsection{Type of AI: Positioning High-risk AI Systems as GPAI Models} The AIA offers very different requirements for AI systems~\citep[Art. 3(1)]{europa} and general-purpose AI (GPAI) models~\citep[Art. 3(63)]{europa}. Our main focus, however, is on avoision tactics that might opportunistically arbitrage the  difference between AI systems and GPAI. As part of the AIA's ``risk-based'' approach, AI initiatives ``are classified into a series of graded `tiers', with proportionately more demanding legal obligations that vary in accordance with the EU’s perceptions of the severity of the risks they pose'' \citep{Smuha2024-jf}. On the one end, ``AI systems posing minimal or no risk are not required to adhere to any of the AIA’s obligations or harmonized standards''~\citep{wachter26limitations}. On the other, the AIA sets a much higher bar for the riskiest class of AI systems --- high-risk AI systems --- than it does the riskiest class of GPAI models--those that pose systemic risk. Accordingly, we can expect that certain AI projects which may bear some level of risk in the eyes of their operators will be positioned as GPAI models rather than AI systems. To illustrate how this behavior could manifest, we present a case study:
 
\paragraph{Case Study: GPAI for education} As detailed in~\citep[Annex III]{europa},  ``education and vocational training'' is considered a high-risk domain. The calculus of regulatory burden between releasing an AI system in the education sector and a GPAI model would then ikely become one between a high-risk AI system and, at most, a GPAI that poses systemic risk. However, because of fewer requirements surrounding GPAI models~\citep{wachter26limitations}, there are incentives for AI systems used in the educational sector to be productionized as GPAI models. By productionizing an AI system as GPAI, we mean that the rhetoric surrounding its release may be that the model or system is a GPAI --- with one of its many uses being in the education sector --- rather than an AI system intended for use in the education domain.\footnote{This may already be happening, with AI technologies described as ``general''~\citep{gemini2023ai} or ``general-purpose''~\citep{anthropic2025aieducation} also being productionized in the education sector~\citep{anthropic2025aieducation,google2025aieducation}.} This shift in framing has the effect of giving firms the same level of access and potentially the same level of impact in the education sector with fewer requirements.

\subsubsection{Categories of Risk: Avoiding Systemic Risk in GPAI}\label{sec:systemicrisk}
With the AIA's risk-based approach to regulation, more requirements accrue as AI models and systems are deemed to carry more risk~\citep{Smuha2024-jf,wachter26limitations}. Positioning AI models and systems in categories of lower risk could then be within firms' self-interest because the firm would have fewer burdensome requirements. For example, GPAI that is categorized as bearing systemic risk incurs additional demands~\citep[Art. 52-53]{europa} such as notification to the EU Commission~\citep[Art. 52.1]{europa}, additional transparency and cybersecurity requirements, increased oversight through the AI office and increased documentation responsibilities~\cite[Art. 55 ]{europa}. A GPAI is considered to bear systemic risk when it meets some of the criteria listed in~\citep[Art. 51]{europa}. Firms may take steps to stay within the boundaries of guidelines listed in ~\citep[Art. 51]{europa} while violating the spirit of the AIA by creating models that might carry as much risk as those just on the other side of the boundary.

\paragraph{Benchmark Shopping and Sandbagging}
As per~\citep[Art. 51, 1(a)]{europa}, a GPAI may be classified as having systemic risk if ``it has high impact capabilities evaluated on the basis of appropriate technical tools and methodologies, including indicators and benchmarks''. Requirements for GPAI with systemic risk could act as counterweight\footnote{For an AI paper to be published, it is typically required that the paper demonstrates significant improvements on various model benchmarks~\cite{lipton2019troubling}. This has led to ``mis-reporting'', or claiming more general reasoning capabilities for their models based on performance on a narrow benchmark~\citep{leech2024questionable} so that research gains traction.} to research incentives to over-sell model performance on benchmarks. Due to additional requirements that accompany the categorization of an AI as GPAI with systemic risk, AI developers may have incentives to selectively evaluate their AI systems on benchmarks, or to ``benchmark shop'', such that evaluations on them do not paint their models as having high-impact capabilities. Benchmark shopping has parallels to other strategies in AI compliance and research that have been discussed in the past like ``audit-washing''\footnote{Audit-washing has been defined as ``acquisition of sustainability or ethical credibility through cosmetic or trivial steps''~\cite{Goodman2022-fu}.} and, perhaps more relatedly,``fairwashing'', defined as ``promoting the false perception that the learning models
used by the company are fair while it might not be so''~\citep{aivodji2019fairwashing}. In particular, \citet{aivodji2019fairwashing} show that it is possible to build a system that systematically searches for fairness metrics that justify a particular decision, even if that decision could be considered unfair.

Another tactic that model developers might pursue to avoid their model's classification as posing systemic risk is ``sandbagging'', where the models themselves are calibrated to ``strategically underperform''~\citep{van2024ai} on benchmarks. As \citet{van2024ai} shows, it is possible to fine-tune GPT-4 and Claude on a synthetic dataset such that prompting the  language models with a ``password'' triggers specific performances given a benchmark. An analogous strategy of installing ``defeat devices'' on motor vehicles, so that emissions are reduced during testing. This strategy was uncovered as part of Volkswagen's emissions scandal~\citep{hotten2015volkswagen}. 

\paragraph{FLOP-Gaming: Small Models and Decentralized Training}

Under the AIA, a GPAI model should be classified as bearing systemic risk, inviting more demanding requirements, if ``the cumulative amount of computation used for [their] training measured in floating point operations is greater than $10^{25}$''~\citep[Art. 51(2)]{europa}. This threshold has been widely criticized as a haphazard proxy for risk level ~\citep{hooker2024limitations} and could be exploited by AI firms through a number of avoisive strategies. As foundational model scaling laws fray at the edges and smaller models prove as powerful as their larger peers \citep{wachter26limitations}, this strategy may provide a way to circumvent additional requirements without any trade-off in terms of model capability. Differently, AI developers may use techniques that distill smaller, sub-threshold models from larger, supra-threshold ones\citep[pg. 698]{wachter26limitations}, a technique that has recently been used to create smaller models that perform nearly as well as their peers on relevant benchmarks ~\cite{DBLP:journals/corr/abs-2402-04616}. Lastly, AI developers may gravitate towards AI development workflows, including but not limited to federated ~\cite{hooker2024limitations} and decentralized learning that, because they spread training across multiple (sometimes inaccessible) devices, render the amount of compute used during training very difficult to measure --- potentially providing developers with a reasonable excuse for not knowing their models triggered the classification ~\citep{skowron2024eleutherai}.

\subsubsection{Categories of Operators: Finger-pointing}

As described in ~\ref{ssec:definingfeatures}, the AI's various operator roles carry different regulatory obligations \citep[Art. 3(8), 25]{europa}. Simply put, AI firms have an incentive to arbitrage between these different operator roles depending on the different requirements associated with the roles. For example, the most onerous are borne by the ``provider'' role, which the AIA defines as anyone that develops an AI model or system (or that has an AI project developed) and places it on the market or puts it into service under their own name or trademark \citep[Art. 3(3)]{europa}. This contrasts with the lighter obligations borne by the role of the deployer, defined as ``a natural or legal person, public authority, agency or other body using an AI system under its authority except where the AI system is used in the course of a personal non-professional activity'' \citep[Art. 3(4)]{europa}. 

Dynamics between providers and deployers might then lead to the off-loading of more burdensome requirements that come with certain role categorizations. The maneuvers we detail bear similarities to those that have been employed to lessen the burden of roles as detailed in the EU General Data Protection Regulation (GDPR)~\cite{europagdpr}. Under the GDPR, technical and organizational forms of ``finger-pointing'' have been leveraged to take advantage of the relational structure of data processors versus data controllers. For instance, Microsoft has deflected responsibility to provide students with information about their data by arguing that the students' school is the controller of the data~\cite{noybmicrosoft2024}. By off-loading data processing and collection onto user devices using privacy-preserving techniques, companies may also avoid the categorization as controllers of user data and have fewer responsibilities regarding that data~\citep{yew2024you,veale2023rights}. In the context of the AIA, the provider is the most burdensome role to take on~\citep{cdtPublicAuthorities}. This creates incentives for operators in other roles to maintain their categorizations and for providers to shed theirs when possible. Below, we consider avoision strategies that operators might employ to achieve this.

\paragraph{Providers: Narrowing Intended Uses}
Providers have an opportunity to transition out of their role when a downstream actor modifies ``the intended purpose of an AI system, including a [GPAI] system''~\cite[Art. 25.1(c)]{europa}. Because providers pass the baton to downstream actors when this occurs~\citep[Art. 25.2]{europa}, there may be incentives to be unimaginative about downstream use cases and to be myopic in the specification of the system's intended purpose. This both lessens the burden in terms of testing and evaluation as part of the risk management system and lowers the barrier for downstream deployers to adopt the categorization of providers and the responsibilities associated.
 
\paragraph{Deployers of High-risk AI Systems: Constraining Technical Changes}
As per \cite[Art. 25]{europa}, one way that a deployer might be subject to the obligations of the provider of a high-risk AI system is through making a substantial modification to a high-risk AI system ``that has already been placed on the market or has already been put into service in such a way that it remains a high-risk AI system pursuant to Article 6''.\footnote{This point is also made as part of \cite[Rec. 84]{europa}: ``This would also be the case [responsibilities of a provider are adopted] if that party makes a substantial modification to a high-risk AI system that ha already been placed on the market or has already been put into service in a way that it remains a high-risk AI system in accordance with this Regulation, or if it modifies the intended purpose of an AI system...''}. For instance, deployers might constrain the technical changes they make to an AI system in order to toe the line of ``substantial modification'' (the point at which provider responsibilities begin). Deployers might decide to prompt-tune the model rather than fine-tune it to argue that the modification of the system should not be considered substantial. Fine-tuning a model means training the last layer of a neural network (NN), and typically guides the model towards specific downstream applications. Prompt-tuning, on the other hand, guides the behavior of models through the use of text prompts, like those entered into ChatGPT. Crucially, fine-tuning typically requires an additional training process and additional training examples~\cite{brown2020language}, while prompt-tuning may only require a few additional examples~\cite{yang2022prompt, brown2020language}. Even when the effect of both changes might be the same, prompt-tuning might give deployers leeway to argue that the model or the system was not changed to the degree that would trigger substantial modification. It has been said that this standard by which deployers shoulder the responsibility of providers is not ``measurable and discernible''~\citep{hacker2024comments}, potentially providing leeway for these strategies to take hold.
\section{Conclusion and Policy Recommendations}
This paper's contributions are two-fold. First, we provide a taxonomy of avoision: an adversarial framework for thinking through the technological and organizational ramifications of a piece of legislation. Secondly, we identify examples of AIA avoision which we then situate within the framework we present.  We take care to limit the scope of the taxonomy and analysis we present. The AIA has yet to be interpreted in courts. Based on how courts interpret the AIA, certain strains of avoision might end up being implausible. Working within the text of the AIA, there are also avoision-adjacent strategies that we do not consider within our taxonomy.\footnote{For example, we do not include avoision-adjacent strategies that appear in certain regulatory processes, like  lobbying processes or  standards-setting processes.} Finally, opportunities to create and exploit loopholes and to engage in avoision may simply represent an inherent feature of written legislation~\citep{katz2023circumvention}. Moreover, exploitable loopholes can sometimes be exactly what generates socially beneficial innovation~\citep{sirman2023loophole}. Our overarching hope is that the framework we present and the strategies we identify can guide policymaking efforts and bolster regulatory scrutiny where applicable.

We conclude our paper with a set of policy recommendations. The AIA has gone into effect, but there remain potential opportunities to mitigate the avoision described in this paper (if so desired). The first of these is the setting of technical standards; the EU Commission has asked the European Committee for Standardization (CEN) and the European Committee for Electrotechnical Standardization (CENELEC) to draft technical standards; AI operators that comply with these standards earn a ``presumption of conformity'' with the AIA \citep{cencenelec, europaDocs}. These ``more concrete'' \citep{DBLP:journals/corr/abs-2208-12645} standards represent an important ``opportunity to push for stronger 'on the ground' protections'' \citep{wachter26limitations}. In addition to technical standard setting, AIA enforcement also represents an opportunity to thwart avoision. The AIA tasks market surveillance authorities in EU member states with enforcing its provisions~\citep[Rec. 153, 156]{europa}; the EU's AI Office also plays a role in enforcing its provisions as regards GPAI models~\citep[Rec. 88]{europa}. The recommendations below consider how these different levers may be used to target the different tiers of avoision.

\paragraph{Avoision Targeting the AIA's Scope}
To deter avoision that attempts to circumvent the AIA’s definitions of AI systems and GPAI models, including by adding ``veneers" of humans, rule-based systems, and traditional software, the EU lawmakers should consider clarifying that these architectures are still within the scope of the AIA.\footnote{A more severe solution, which may lack public support, would be to broaden the scope of the AIA by ``changing the name from `Artificial Intelligence Act' to `Algorithms Act' or `Software Act.' " and explicitly include rules-based and traditional software systems in its scope \citep{smuha2021eu}.} This can be done in future amendments to the AIA or, perhaps, in the forthcoming technical specifications. To deter avoision aimed at obfuscating AI systems and models' connections to the EU market, including by turning a blind eye to illegitimate use of their products or adding human and software ``veneers'', EU lawmakers should consider clarifying that these behaviors are still within the scope of the AIA. 

\paragraph{Avoision Targeting the AIA's Exemptions}
When it comes to avoison designed to exploit the AIA's exemptions, forthcoming technical specifications through the standards-setting process can play an important role. In particular, these mitigations can focus on adding criteria to the exemptions that keep them aligned with their original intentions.  As part of Section~\ref{sec:opensourceexploit}, we consider how firms may hide behind the ambiguity in ``released under free and open-source licenses'' to release models that are prohibitively expensive or not compliant to run. Technical standards could specify exactly how AI should be open-sourced (and what should be open sourced). Standards-setting may present an opportunity to refine the characteristics of open-source licenses that should allow developers to fall under open-source exemptions.
 
\paragraph{Avoision Targeting the AIA's Consequential Categories}

A shared vulnerability enabling different avoision strategies that target the AIA's consequential categories is how technical protections and the rhetoric that surrounds them can be used to shield firms from liability. Independent assessments and standards-setting can help to scrutinize technical protections and language against how they might serve avoisive behaviors. Echoing the recommendations of some scholars \citep{wachter26limitations, smuha2021eu}, the EU might consider requiring that risk assessments be conducted by independent third parties. Some of the flexibility firms have in arbitraging between different categories of AI as we describe in Section~\ref{sec:categorizations} arises from the AIA's self-grading loophole. While~\citep[Recital 125]{europa} eventually envisions independent third-party conformity assessments~\citep{wachter26limitations}, it also notes that, given the current level of experience of certifiers, the scope of third-party assessments should be limited. Standards-setting can also play an important role in clarifying both how FLOPs should be calculated when models are trained in a decentralized fashion, as well as the conditions under which decentralized training actually confers security protections. Given that benchmarks have been used whether or not they confer meaningful information about model performance~\citep{raji2021ai,leech2024questionable}, standards-setting can also play a role in addressing avoision tactics by guiding firms to benchmarks under different contexts that meaningfully assess an AI model or system's performance under a given context.

% This section has a special environment:
% \begin{verbatim}
\begin{acks}
We thank the reviewers for their helpful insight. R.Y. is funded through support from the MacArthur and Ford Foundation. She dedicates this work to her late research mentor, Lawrence E McCray (LMc). 
\end{acks}
\bibliographystyle{ACM-Reference-Format}
\bibliography{sample-base}

%%
%% The next two lines define the bibliography style to be used, and
%% the bibliography file.

%%
%% If your work has an appendix, this is the place to put it.
% \appendix

% \section{Research Methods}

% \subsection{Part One}

% Lorem ipsum dolor sit amet, consectetur adipiscing elit. Morbi
% malesuada, quam in pulvinar varius, metus nunc fermentum urna, id
% sollicitudin purus odio sit amet enim. Aliquam ullamcorper eu ipsum
% vel mollis. Curabitur quis dictum nisl. Phasellus vel semper risus, et
% lacinia dolor. Integer ultricies commodo sem nec semper.

% \subsection{Part Two}

% Etiam commodo feugiat nisl pulvinar pellentesque. Etiam auctor sodales
% ligula, non varius nibh pulvinar semper. Suspendisse nec lectus non
% ipsum convallis congue hendrerit vitae sapien. Donec at laoreet
% eros. Vivamus non purus placerat, scelerisque diam eu, cursus
% ante. Etiam aliquam tortor auctor efficitur mattis.

% \section{Online Resources}

% Nam id fermentum dui. Suspendisse sagittis tortor a nulla mollis, in
% pulvinar ex pretium. Sed interdum orci quis metus euismod, et sagittis
% enim maximus. Vestibulum gravida massa ut felis suscipit
% congue. Quisque mattis elit a risus ultrices commodo venenatis eget
% dui. Etiam sagittis eleifend elementum.

% Nam interdum magna at lectus dignissim, ac dignissim lorem
% rhoncus. Maecenas eu arcu ac neque placerat aliquam. Nunc pulvinar
% massa et mattis lacinia.

\end{document}